\RequirePackage{fix-cm}
\documentclass[smallextended]{svjour3}       

\smartqed  
\usepackage{graphicx}
\usepackage{mathptmx}      
\usepackage{amsmath}
\usepackage{fancyvrb}
\usepackage{verbatim}
\usepackage{amssymb}
\PassOptionsToPackage{hyphens}{url}

\usepackage{url}
\usepackage[breaklinks]{hyperref}

\usepackage{breakurl}

\DefineVerbatimEnvironment{rlispverb}{Verbatim}{
framesep=2mm,xleftmargin=\parindent}

\newcommand*{\fd}
[2]{\mathchoice{\frac{\delta#1}{\delta#2}}
  {\delta#1/\delta#2}{\delta#1/\delta#2}
  {\delta#1/\delta#2}}
\newcommand{\ddx}[1]{\partial_x^{#1}}
\begin{document}

\title{WDVV equations: symbolic computations of Hamiltonian operators}

\author{Jakub Vašíček        \and
        Raffaele Vitolo 
}

\institute{J. Vašíček \at
            Mathematical Institute of the Silesian University,\\
			Silesian University, Opava, Czech Republic\\
            \email{jakub.vasicek@math.slu.cz}
\and
           R. Vitolo \at
            Department of Mathematics and Physics  ``E. De Giorgi'' ,\\
			University of Salento, Lecce, Italy\\
			\email{raffaele.vitolo@unisalento.it}  
}

\date{
    \framebox{
\begin{minipage}{8cm}
  \begin{center}
  \footnotesize  Published in\\
  \footnotesize 
Applicable Algebra in Engineering, Communication and Computing
  \\
  DOI: \url{https://doi.org/}\\
  arXiv: \url{https://arxiv.org/abs/}\\[3mm]
  {\itshape Dedicated to the memory of Vladimir Gerdt (1947--2021)}
\end{center}
\end{minipage}
}
}

\maketitle

\begin{abstract}
  We describe software for symbolic computations that we developed in order to
  find Hamiltonian operators for Witten--Dijkgraaf--Verlinde--Verlinde (WDVV)
  equations, and verify their compatibility. The computation involves nonlocal
  (integro-differential) operators, for which specific canonical forms and
  algorithms have been used.

\keywords{WDVV equations, Hamiltonian operators, Schouten bracket, symbolic
  computations, integro-differential operators, weakly nonlocal operators.}
  
\subclass{37K10, 53D45, 68W30.}
\end{abstract}

\section{Introduction}
\label{intro}
The Witten-Dijkgraaf-Verlinde-Verlinde (WDVV) equations, also known as
associativity equations, originated in $2D$-topological field theory, but they
acquired mathematical interest from the work of B. Dubrovin \cite{D96}, who
showed the deep connections between solutions of WDVV equations and integrable
systems of PDEs, in particular, bi-Hamiltonian hierarchies.

Recently \cite{VV21}, we proved that, in low dimensions, the
WDVV equations themselves are bi-Hamiltonian systems of PDEs. The proof
required sophisticated symbolic computational tools that have been only
recently made available \cite{CLV19,m.20:_weakl_poiss}.

In this paper, we would like to describe the algorithms and symbolic software
that we developed in order to carry out the computations in \cite{VV21}. To do
that, let us first describe the mathematical problem that we solved. Following
\cite{dubrovin06:_encyc_mathem_physic}, in $\mathbb{R}^N$ we are to find a
function $F=F(t^1,\ldots,t^N)$ such that\label{page:introduction}
\begin{enumerate}
\item
  \begin{sloppypar}$\displaystyle \frac{\partial^3 F}{\partial t^1\partial
      t^\alpha\partial t^\beta} = \eta_{\alpha\beta}$ is a constant symmetric
    nondegenerate matrix;
  \end{sloppypar}
\item
  $c^\gamma_{\alpha\beta} = \eta^{\gamma\epsilon} \displaystyle
  \frac{\partial^3 F}{\partial t^\epsilon\partial t^\alpha\partial t^\beta}$
  are the structure constants of an associative algebra;\label{item:2}
\item $F$ is quasihomogeneous:
  $F(c^{d_1}t^1,\ldots,c^{d_N}t^N) = c^{d_F}F(t^1,\ldots,t^N)$.\label{item:1}
\end{enumerate}
The conditions of associativity, or WDVV equations, are the system of PDEs
\begin{equation}\label{eq:5}
  \eta ^{\mu \lambda }\frac{\partial ^{3}F}{\partial t^{\lambda }\partial
    t^{\alpha }\partial t^{\beta }}\frac{\partial ^{3}F}{\partial t^{\nu
    }\partial t^{\mu }\partial t^{\gamma }}=\eta ^{\mu \lambda }\frac{\partial
    ^{3}F}{\partial t^{\nu }\partial t^{\alpha }\partial t^{\mu }}\frac{\partial
    ^{3}F}{\partial t^{\lambda }\partial t^{\beta }\partial t^{\gamma }}.
\end{equation}
Item $1$ implies that $F$ can be rewritten as
\begin{equation}\label{eq:71}
  F=\frac{1}{6}\eta_{11}(t^1)^3 +
  \frac{1}{2}\sum_{k>1}\eta_{1k}t^k(t^1)^2 + \frac{1}{2}\sum_{k,s>1}
  \eta_{sk}t^st^kt^1 + f(t^2,\ldots,t^N),
\end{equation}
up to a second degree polynomial of the field variables.  This implies that the
WDVV system is an overdetermined system in one unknown function $f$ of $N-1$
independent variables.

A technique introduced in \cite{mokhov95:_sympl_poiss} (see also
\cite{ferapontov96:_hamil}) allows to rewrite the WDVV system as $N-2$
commuting quasilinear first-order systems of conservation laws of the form
\begin{equation}
  \label{eq:7}
  u^i_t = (V^i(\mathbf{u}))_x,\qquad i=1,\ldots,n.
\end{equation}
For example, if $N=3$ and
\begin{equation}\label{eq:8}
  \eta_{\alpha\beta}= \delta_{\alpha+\beta,N+1}=
  \begin{pmatrix}
    0 & 0 & 1
    \\
    0 & 1 & 0
    \\
    1 & 0 & 0
  \end{pmatrix},
\end{equation}
there is only one WDVV equation:
\begin{equation}\label{eq:13}
  f_{ttt}=f_{xxt}^2 - f_{xxx}f_{xtt}.
\end{equation}
Introducing the new coordinates $a=f_{xxx}$, $b=f_{xxt}$,
$c=f_{xtt}$ we have the compatibility conditions
\begin{equation}\label{eq:6}
  \left\{
    \begin{array}{l}
      a_t = b_x,\\ b_t = c_x,\\ c_t = (b^2 - ac)_x
    \end{array}
  \right.
\end{equation}
The above representation allowed to look for a bi-Hamiltonian formalism. That
was found in \cite{FGMN97}: the right-hand side of \eqref{eq:6} was rewritten
in two ways
\begin{equation}\label{eq:16}
  u_{t}^{i}=A_{1}^{ij}\fd{H_2}{u^j}=A_{2}^{ij}\fd{H_{1}}{u^j},
\end{equation}
where $\fd{}{u^i}$ stands for the variational derivative, $H_1$ and $H_2$ are
two Hamiltonian densities and $A_{1}$, $A_{2}$ are two compatible local
Hamiltonian operators.  Here, `compatible' means that the pencil
$A_1+\lambda A_2$ is a Hamiltonian operator for every value of the parameter
$\lambda$. We recall that the Hamiltonian property for a differential operator
is equivalent to differential conditions on its coefficients, to be discussed
later. Such a property is crucial for the integrability of the system of PDEs.

Only a few other examples of bi-Hamiltonian WDVV systems were known until
recently.  In \cite{VV21} we were able to prove that, for
$N=3$, all WDVV systems admit a bi-Hamiltonian formulation, with Hamiltonian
operators of the form
\begin{align}
    A_1^{ij} =& g^{ij}\ddx{} + \Gamma^{ij}_{k}u^k_x + \alpha V^i_qu^q_x\ddx{-1}
    V^j_pu^p_x\notag
    \\ \label{eq:9}
    & +\beta\left( V^i_qu^q_x\ddx{-1}u^j_x + u^i_x\ddx{-1}V^j_qu^q_x \right)
     +\gamma u^i_x\ddx{-1}u^j_x,
  \\
  \label{eq:10}
  A_2 =& \partial_x(h^{ij}\partial_x + c^{ij}_k u^k_x)\partial_x,
\end{align}
with suitable Hamiltonian densities $H_1$ and $H_2$. We observe that $A_1$ is a
nonlocal first-order homogeneous Hamiltonian operator of Ferapontov type
\cite{F95:_nl_ho}. That is a member of a more general family of nonlocal
operators, namely weakly nonlocal operators, that is discussed in
\cite{CLV19}. In this context, `nonlocal' means integro-differential
operator (derivatives are taken with respect to the variable $x$).
$A_2$ is a local third-order homogeneous Hamiltonian operator in canonical form
\cite{FPV14,FPV16}. See more on these operators in next Section.

One of the computations that is needed in order to prove the bi-Hamiltonian
property is the compatibility of the operators. That is made difficult by the
fact that one of the operators is weakly-nonlocal (the Ferapontov operator
$A_1$). The computation has been made possible by the recent computer algebra
packages developed in \cite{m.20:_weakl_poiss}.

Here we will describe how we used the Reduce computer algebra system, see
\url{https://reduce-algebra.sourceforge.io/}, and its CDE package in particular
(see \cite{KVV17} for a general description and \cite{m.20:_weakl_poiss} for
calculations with weakly-nonlocal operators) in the nontrivial calculations
that led to the results presented in \cite{VV21}. We also
checked some of the computations in Maple using the package
\texttt{jacobi.mpl} from \cite{m.20:_weakl_poiss}. Maple was also used in a
significant part of the computation in the case $N=5$.

All files that are described in the text can be found in a \texttt{GitHub}
repository. The files are also downloadable as a single \texttt{.zip} file, see
\cite{vasicekvitolo21:_wdvv_symbcomp}.

\section{Computing the first-order Hamiltonian operator \texorpdfstring{$A_1$}{A1}}
\label{sec:3}

Below we will focus on finding a first-order, weakly nonlocal Hamiltonian
operator of Ferapontov type.

In the expression~\eqref{eq:9} $(g^{ij})$ is a non-degenerate matrix of
functions of the field variables $(u^i)$, whose inverse $(g_{ij})$ can be
interpreted as a covariant $2$-tensor. The matrix $V^i_q$ is just the Jacobian
of the vector function $(V^i)$ of the fluxes of the system~\eqref{eq:7}, and
$\alpha$, $\beta$, $\gamma$ are three constants.

The Hamiltonian property of $A_1$ is equivalent to the following conditions
\cite{F95:_nl_ho}:
\begin{itemize}
\item $g^{ij}$ is symmetric;
\item $\Gamma^{i}_{jk} = - g_{js}\Gamma^{si}_k$ are the 
Christoffel symbols of $g_{ij}$, regarded as a metric;
\item the identities:
\end{itemize}
\begin{subequations}\label{eq:4}
  \begin{align}\label{eq:12}
    &g^{ik}V^j_k = g^{jk}V^i_k,
    \\ \label{eq:14}
    &\nabla_kV^i_j = \nabla_j V^i_k,
    \\ \label{eq:17}
      \begin{split}
    &R^{ij}_{kl} = \alpha\big( V^i_k V^j_l - V^i_l V^j_k \big) \\
    &\hphantom{R^{ij}_{kl} =}
    +\beta\big( V^i_k \delta^j_l - V^j_k \delta^i_l - V^i_l \delta^j_k +
    V^j_l \delta^i_k \big) +\gamma(\delta^i_k\delta^j_l -
    \delta^i_l\delta^j_k)
  \end{split}
  \end{align}
\end{subequations}
hold, where $\nabla$ denotes the covariant derivative with respect to the
Levi-Civita connection of $g_{ij}$ and $R^{ij}_{kl}=g^{is}R^j_{skl}$ is the
Riemannian curvature tensor. The above conditions also imply that the operator
is a Hamiltonian operator for the system of PDEs \eqref{eq:7}
\cite{tsarev85:_poiss_hamil,vergallo20:_homog_hamil}.

Thus, finding an operator \eqref{eq:9} actually reduces to finding the
constants $\alpha, \beta$ and $\gamma$ and the metric $g^{ij}$. However,
finding the metric in general is by no means an easy computational task. To
this end, we shall use a theorem from \cite{bogoyavlenskij96:_neces_hamil}
which states that the metric of a first-order Hamiltonian operator for a
non-diagonalizable quasilinear first-order system in $n=3$ unknown functions is
proportional to a contraction of the square of the Haantjes tensor:
\begin{equation}
  \label{eq:62}
  g_{ij} = f\, H^\alpha_{i\beta}H^{\beta}_{j\alpha}=f\,H_{ij},
  \qquad f=f(\mathbf{u}).
\end{equation}
We recall that the Haantjes tensor is obtained from $(V^i)$ and its
derivatives by means of a straightforward formula (see
\cite{VV21}).

After this simplification we are left with a much less demanding computational
problem -- we need to find one unknown function $f(\mathbf{u})$ and the above
constants. To this end, we will use a computer algebra system to determine our
unknowns.

\subsection{The computation}
\label{sec:3.2}

The computation of a nonlocal first-order operator is divided into three
smaller steps. Firstly, we will write down the candidate metric
$g_{ij} = f\,H_{ij}$ and check~\eqref{eq:12}. Next we will solve the equation
\eqref{eq:14} which will yield an explicit form of the metric $g^{ij}$ and as a
final step, we find the constants $\alpha, \beta$ and $\gamma$ after plugging
the found metric into the condition \eqref{eq:17}. We will concentrate on a
case from Mokhov-Pavlenko's classification~\cite{mokhov18:_class_hamil}, where
the determining matrix $\eta$ has the following form:
\begin{equation}
  \eta^4=
  \begin{pmatrix}
    1 & 0 & 0 \\
    0 & \lambda & 0 \\
    0 & 0 & \mu %
  \end{pmatrix},\ \lambda^2=1, \ \mu^2=1, 
\end{equation}
since it is not as computationally demanding and can be solved in full
generality with respect to the constants $\lambda,\mu$. The WDVV equation takes
the form:
\begin{equation}
 \mu f_{xxt}^2-\mu f_{xxx}f_{xtt}+\lambda f_{xtt}^2-\lambda f_{xxt}f_{ttt} -1= 0.
\end{equation}
After introducing the coordinates $u^1=f_{xxx}$, $u^2=f_{xxt}$, $u^3=f_{xtt}$
we obtain the first-order WDVV system in conservative form:
\begin{equation}
  \label{eq:83}
  \begin{split}
    & u^1_t = u^2_x,\\
    & u^2_t = u^3_x,\\
    & u^3_t = \left(\frac{\mu ((u^2)^2- u^1u^3)+\lambda (u^3)^2-1}
      {\lambda u^2}\right)_x.
  \end{split}
\end{equation}
All files used for computation are available at a \texttt{GitHub} repository \cite{vasicekvitolo21:_wdvv_symbcomp}.

\subsubsection{The first step}
\label{sec:3.2.1}
The following computation can be found in the file
\texttt{WDVV-3c-Eta4\textbackslash dne3\_lho2.red}. In Reduce, we start by
loading the package \texttt{cde} and initialization of our environment by the
following sequence:
\begin{rlispverb}
indep_var:= {x};
dep_var:= {u1,u2,u3};
total_order:= 8;
cde({indep_var,dep_var,{},total_order},{});
\end{rlispverb}
followed by initialization of the right hand-side of the system:
\begin{rlispverb}
de:={u2_x, u3_x, 
	td((mu*u2^2 - mu*u1*u3 + lam*u3^2 - 1)/(lam*u2),x)};
\end{rlispverb}
where the last element is just the WDVV equation.
Next, define the velocity matrix $V^i_j$ (and its operator) of our system:
\begin{rlispverb}
nc:=length(dep_var);
matrix av(nc,nc);
for i:=1:nc do
  for j:=1:nc do
    av(i,j):=df(part(de,i),part(ford_var,j));
    
operator avt;
for i:=1:ncomp do for j:=1:ncomp do
  avt(i,j):=av(i,j);    
\end{rlispverb}

We also need to load a \texttt{riemann4.red} library in order to generate the
Nijenhuis and Haantjes tensors by the commands:
\begin{rlispverb}
generate_all_nt(avt,dep_var);
generate_all_ht(avt,dep_var);
\end{rlispverb}
And finally the contraction of the square of Haantjes tensor is obtained by:
\begin{rlispverb}
temphtc:={};
matrix hmet(nc,nc);
for i:=1:nc do
  for j:=1:nc do
    hmet(i,j):=
    <<
      temphtc:=for k:=1:nc join for h:=1:nc collect
      riemann_list2ids({ht_,k,i,h})*riemann_list2ids({ht_,h,j,k});
      part(temphtc,0):=plus
    >>;
\end{rlispverb}
As a last step we check the symmetry condition $H_{ih}V^h_j = H_{jh}V^h_i$:  
\begin{rlispverb}
for i:=1:nc do for j:=i+1:nc do
<<
  templhs:=(for h:=1:nc sum hmet(i,h)*av(h,j));
  temprhs:=(for h:=1:nc sum hmet(j,h)*av(h,i));
  write templhs - temprhs
>>;
\end{rlispverb}
and save the result \texttt{hmet} for later computations. We remark that, by
construction, \texttt{hmet} is a rational function of the field variables
\texttt{u1, u2, u3} and the parameters \texttt{lam, mu}.

\subsubsection{The second step}
\label{sec:3.2.2}

The following computation can be found in the file
\texttt{WDVV-3c-Eta4\textbackslash dne3\_lho3.red}. We, again, start by
initializing the environment as in previous step. After constructing the
velocity matrix $V^i_j$ and loading the previous result we proceed to defining
the metric $g_{ij}$ as a functional factor of the contraction of the square of
the Haantjes tensor:
\begin{rlispverb}
for each el in dep_var do depend f,el;
gl1:=f*hmet;
\end{rlispverb}
Besides the metric itself, we will need its inverse $g^{ij}$ and Christoffel
symbols of the third kind $\Gamma^{ij}_k$ (which are constructed inside the
package \texttt{riemann3.red} by means of Christoffel symbols of the first and
second kind):
\begin{rlispverb}
gu1:=gl1^(-1);
operator gl1_op,gu1_op;
for i:=1:nc do for j:=1:nc do gl1_op(i,j):=gl1(i,j);
for i:=1:nc do for j:=1:nc do gu1_op(i,j):=gu1(i,j);

generate_all_chr1(gl1_op,chr1_,dep_var);
generate_all_chr2(gl1_op,gu1_op,chr1_,chr2_,dep_var);
generate_all_chr3(gl1_op,gu1_op,chr2_,chr3_,dep_var);

operator gamma_hi;
for i:=1:nc do for j:=1:nc do for k:=1:nc do
  gamma_hi(i,j,k):=mk_chr3(chr3_,i,j,k);
\end{rlispverb}
To find the coefficients of the metric we will use the condition \eqref{eq:14}
which is in our case equivalent to:
\begin{equation*}
V^k_j \Gamma^{si}_k = V^s_k \Gamma^{ki}_j.
\end{equation*}
Hence, we first construct the left and right hand-side respectively:
\begin{rlispverb}
operator ag1;
for all s,i,j let ag1(s,i,j)=
  (for k:=1:nc sum av(k,j)*gamma_hi(s,i,k));
operator ag2;
for all s,i,j let ag2(s,i,j)=
  (for k:=1:nc sum av(s,k)*gamma_hi(k,i,j));
\end{rlispverb}

and we assemble the system:

\begin{rlispverb}
total_eq:=
  for i:=1:nc join
    for j:=1:nc join
      for s:=1:nc collect
        ag1(s,i,j) - ag2(s,i,j);
\end{rlispverb}

Using the package \texttt{crack}, a solver of overdetermined systems of PDEs
\cite{WB,WB95}, we obtain a solution for our unknown function $f(\mathbf{u})$
by the following sequence:
\begin{rlispverb}
split_vars:=cde_difflist(all_parametric_der,dep_var);

splitvars_total_eq:=splitvars_list(total_eq,split_vars);

unk:={f};

load_package crack;
lisp(max_gc_counter:=10000000000);
crack_results:=crack(splitvars_total_eq,{},unk,{});

sol_unk:=second first crack_results;
\end{rlispverb}
Then by a mere substitution, have the explicit form of the metric $g$:
\begin{rlispverb}
gl1:=sub(sol_unk,gl1);
gu1:=sub(sol_unk,gu1);
\end{rlispverb}
We again export this result in order to finish the final step - finding the
nonlocal part of the first-order operator.

\subsubsection{The third step}
\label{sec:3.3.2}

The following computation can be found in the file
\texttt{WDVV-3c-Eta4\textbackslash dne3\_lho4.red}. At last, we need to find
the nonlocal tail of the operator $A_1$. We simply need to plug everything we
found into the condition \eqref{eq:5} and see for which values of constants
$\alpha, \beta, \gamma$ it is satisfied.

As usual, we initialize the environment. As an addition to that we load the
results from the previous file and then construct the metric and generate all
Christoffel symbols $\Gamma^{ij}_k$:
\begin{rlispverb}
gl1:=gu1^(-1);

operator gl1_op,gu1_op;
for i:=1:nc do for j:=1:nc do gl1_op(i,j):=gl1(i,j);
for i:=1:nc do for j:=1:nc do gu1_op(i,j):=gu1(i,j);

generate_all_chr1(gl1_op,chr1_,dep_var);
generate_all_chr2(gl1_op,gu1_op,chr1_,chr2_,dep_var);
generate_all_chr3(gl1_op,gu1_op,chr2_,chr3_,dep_var);
\end{rlispverb}
Let us also define the identity matrix which will represent the Kronecker
$\delta^i_j$:
\begin{rlispverb}
idm:=mat((1,0,0),(0,1,0),(0,0,1));
\end{rlispverb}
and finally assemble the condition \eqref{eq:17}:
\begin{rlispverb}
eq_curv:=for i:=1:nc join
for j:=1:nc join
  for k:=1:nc join
    for h:=1:nc collect
      riem3(gl1_op,gu1_op,chr2_,i,j,k,h,dep_var) - 
      alp*(av(i,k)*av(j,h) - av(i,h)*av(j,k)) - 
      bet*(av(i,k)*idm(j,h) - av(j,k)*idm(i,h) - 
        av(i,h)*idm(j,k) + av(j,h)*idm(i,k)) - 
      gam*(idm(i,k)*idm(j,h) - idm(i,h)*idm(j,k));
\end{rlispverb}
The system is a rational expression that should vanish for any value of the
field variables \texttt{u1, u2, u3}. That yields an overdetermined linear
system on the coefficients \texttt{alpha, beta, gamma} that can be solved only
if \texttt{alp=mu, bet=0, gam=lam}. The substitution:
\begin{rlispverb}
A:=sub({
  alp=mu,
  bet=0,
  gam=lam
  },eq_curv);
write eq_curv;
\end{rlispverb}
yields a list of non-zero equations in terms of the parameters \texttt{mu,
  lam}. However, we should check only for all the possible combinations of the
values of constants $\mu, \lambda$:
\begin{rlispverb}
sub({mu=1,lam=1},A);
sub({mu=-1,lam=1},A);
sub({mu=1,lam=-1},A);
sub({mu=-1,lam=-1},A);
\end{rlispverb}
which finally yields four lists of zeroes.

The metric $g^{ij}$ that defines $A_1$, according with the expression
\eqref{eq:9}, is:
\begin{equation}
  g^{ij}= \frac{\lambda}{\mu}
  \begin{pmatrix}
    -a^2\mu-b^2\lambda-4\mu\lambda & -b( a\mu+c\lambda) &
    -b^2\mu - c^2\lambda - 1 \cr 
    -b( a\mu+c\lambda) & -b^2\mu - c^2\lambda - 1 &
    \frac{c( ac\mu -2b^2\mu -c^2\lambda + 1)}{b} \cr 
    -b^2\mu - c^2\lambda - 1 & \frac{c(ac\mu -2b^2\mu -c^2\lambda + 1)}{b}
    & \frac{\delta}{b^2}
  \end{pmatrix},
\end{equation}
where $\delta= 2ab^2c\lambda-a^2c^2\lambda +2ac^3\mu -2ac\mu\lambda
-b^4\lambda -3b^2c^2\mu -2b^2\mu\lambda -c^4\lambda +2c^2 - \lambda$
and the values of constants in \eqref{eq:9} are
$\alpha=\mu,\beta=0, \gamma=\lambda$, with $\lambda,\mu = \pm 1$.

\section{Computing the third-order Hamiltonian operator
  \texorpdfstring{$A_2$}{A2}}
\label{sec:third}

To find a bi-Hamiltonian structure for the system~\eqref{eq:7} we need a second
Hamiltonian operator compatible with $A_1$. It was proved in
\cite{VV21} that in the case $N=3$, $4$ and $5$ the WDVV
systems admit a third-order homogeneous Hamiltonian operator in the canonical
form
\begin{equation}\label{eq:3}
  A_2 = \partial_x(h^{ij}\partial_x + c^{ij}_ku^k_x)\partial_x
\end{equation}
(we always require $\det\,(h^{ij})\neq 0$; $(h_{ij})$ denotes the inverse
matrix). We recall that the Hamiltonian property of $A_2$ is equivalent to the
conditions
\begin{subequations}
  \begin{align}\label{eq:1}
    & c_{ijk}=\frac{1}{3}(h_{ik,j} - h_{ij,k}).
    \\\label{eq:2}
    & h_{mk,s} + h_{ks,m} + h_{ms,k}=0,
    \\\label{eq:11}
    & c_{msk,l}= - h^{pq}c_{pml}c_{qsk}.
\end{align}
\end{subequations}
and that the equation~\eqref{eq:2} implies that $h_{ij}$ is a \emph{Monge
  metric} \cite{FPV14,FPV16}. For computational purposes, that implies that
the coefficients $h_{ij}$ are quadratic polynomials of the field variables.

Then, it is also known~\cite{FPV17:_system_cl} that the conditions under which
a third-order Hamiltonian operator is the Hamiltonian operator for a
quasi-linear system of first-order conservation laws are:
\begin{subequations}
  \label{eq:3rdOrdCon}
  \begin{align}\label{e1}
    & h_{im}V^{m}_{j}=h_{jm}V^m_{i},\\
    \label{e3}
    & c_{mkl}V^m_{i}+c_{mik}V^m_{l}+c_{mli}V^m_{k}=0,\\
    \label{e2}
    &h_{ks}V^k_{ij}=c_{smj}V^m_{i} + c_{smi}V^{m}_{j},
  \end{align}
\end{subequations}
Thus, with respect to the problem of finding a third-order operator for WDVV
systems, knowing the vector of fluxes $(V^i(\mathbf{u}))$ reduces the above
system of equations to a linear system of algebraic equations in the
coefficients of the quadratic polynomials $h_{ij}$ that can be solved in short
time for small values of $N$.

For the particular example that we discussed in the previous section (system
\eqref{eq:83}) we can easily solve the system \eqref{eq:3rdOrdCon} and verify
that the unique solution $h_{ij}$ that we get fulfills the conditions
\eqref{eq:2} and \eqref{eq:11}. This implies that $h_{ij}$ generates a
third-order homogeneous Hamiltonian operator  \eqref{eq:3} using the formula
\eqref{eq:1}. We have the expression:
\begin{equation}
  h_{ij}=
  \begin{pmatrix}
    b^2+\mu & b\mu(\lambda c-\mu a) & -\mu\lambda b^2 \\
    b\mu(\lambda c-\mu a) & \lambda+a^2-\lambda c(2\mu a-\lambda c) &
    \lambda b(\mu a-\lambda c) \\
   -\mu\lambda b^2 & \lambda b(\mu a-\lambda c)) & b^2 \\
  \end{pmatrix}.
\end{equation}

As this computation is not using any special package, there is no need to go
into details in the code itself.  The computation is provided in the file
\texttt{WDVV-3c-Eta4\textbackslash \\ wdvv\_3ord\_op\_eta4.red}

\section{Compatibility of the Hamiltonian operators \texorpdfstring{$A_1$}{A1},
  \texorpdfstring{$A_2$}{A2}}
\label{sec:4}

A \emph{bi-Hamiltonian} system is just a system of PDEs which is Hamiltonian
with respect to two Hamiltonian operators, $A_1$ and $A_2$ in our case, which
are compatible. The compatibility condition can be either checked by means of
the requirement that the pencil $A_1+\lambda A_2$ is a Hamiltonian operator for
every value of the parameter, or, equivalently, by computing the Schouten
bracket $[A_1,A_2]$ (see \cite{CLV19}).

However, tensorial conditions which are equivalent to the compatibility
condition of $A_1$ and $A_2$ as in our case are not available. So, the
calculation should be made through the definition, which makes a pen-and-paper
approach to the problem almost completely unfeasible. This is also due to the
fact that an algorithm to do the calculation in the nonlocal case was missing
until recently. A computational algorithm \cite{CLV19} and software packages
\cite{m.20:_weakl_poiss} became available a short time ago. Through this
software we checked the compatibility of the operators $A_1$ and $A_2$.

All of the computations described below are contained in the files \linebreak
\texttt{wdvv\_comp\_nl1\_eta4.red} and \texttt{wdvv\_comp\_nl2\_eta4.red}. For
convenience they are split into two separate files where in the first one we
check the bracket [$A_1,A_1$] and then in the second file we check the
compatibility condition [$A_1,A_2$]. Let us focus on the second computation.

We start as usual by initializing the environment:
\begin{rlispverb}
indep_var:= {x};
dep_var:= {u1,u2,u3};
total_order:= 8;
cde({indep_var,dep_var,{},total_order},{});

de:={u2_x, u3_x, 
	td((mu*u2^2 - mu*u1*u3 + lam*u3^2 - 1)/(lam*u2),x)};
\end{rlispverb}

From now on, we initialize the Ferapontov operator $A_1$. The metric of the
first-order operator is loaded from another file:
\begin{rlispverb}
in "dne3_lho3_res.red";
gu1:=gl1**(-1);
\end{rlispverb}
To construct the local part we further need to initialize the Christoffel
symbols $\Gamma^{ij}_k$ in the same way as before. We construct the local part
of the operator by the following lines:
\begin{rlispverb}
operator b;
for i:=1:nc do for j:=1:nc do
  b(i,j):=for k:=1:nc sum mk_chr3(chr3_,i,j,k)*part(dv1,k);

mk_cdiffop(ham1_l,1,{3},3);
for all i,j,psi let ham1_l(i,j,psi)=
  gu1(i,j)*td(psi,x) + b(i,j)*psi;
\end{rlispverb}
where $b$ represents the contraction $\Gamma^{ij}_k(\mathbf{u})u^k_x$.

The other, nonlocal, part of our operator consists of a nonlocal tail which
is defined by the previously found values of constants $\alpha, \beta, \gamma$:
\begin{rlispverb}
mk_wnlop(c,w,2);
c(1,1):= mu;
c(2,2):= rho;
c(1,2):=0;
c(2,1):=0;
for i:=1:nc do w(i,1):=(for j:=1:nc sum av(i,j)*part(dv1,j));
w(1,2):=u1_x;
w(2,2):=u2_x;
w(3,2):=u3_x;
\end{rlispverb}
Here the constants are represented by a symmetric matrix $c^{ij}$; $w^k_i$ on
the other hand represents the nonlocal terms.

Now, we import the metric (in lower indices) of the third-order operator,
stored in \texttt{gl3}, and we generate the constants
$c_{ijk}= \frac{1}{3}(h_{ki,j} - h_{ji,k})$:
\begin{rlispverb}
operator c_lo;
for i:=1:nc do for j:=1:nc do for k:=1:nc do
  c_lo(i,j,k):= (1/3)*(df(gl3(k,i),part(dep_var_equ,j))
  		 - df(gl3(j,i),part(dep_var_equ,k)));
\end{rlispverb}
then we raise two indices:
\begin{rlispverb}
templist:={};
operator c_hi;
for i:=1:nc do for j:=1:nc do for k:=1:nc do
   c_hi(i,j,k):=
    <<
     templist:=
       for m:=1:nc join for n:=1:nc collect
         gu3(n,i)*gu3(m,j)*c_lo(m,n,k);
     templist:=part(templist,0):=plus
    >>;
\end{rlispverb}
and finally we need the contraction $c^{ij}_ku^k_x$:
\begin{rlispverb}
operator c_hi_con;
for i:=1:nc do for j:=1:nc do
  c_hi_con(i,j):=
   <<
    templist:=for k:=1:nc collect 
    	c_hi(i,j,k)*mkid(part(dep_var_equ,k),!_x);
    templist:=part(templist,0):=plus
   >>;
\end{rlispverb}
We assemble the local part:
\begin{rlispverb}
mk_cdiffop(ham2_l,1,{3},3);
for all i,j,psi let ham2_l(i,j,psi) =
td( gu3(i,j)*td(psi,x,2)+c_hi_con(i,j)*td(psi,x), x);
\end{rlispverb}
and even though there is no nonlocal part, we need to introduce a zero tail:
\begin{rlispverb}
mk_wnlop(d,z,1);
d(1,1):=0;
for i:=1:nc do z(i,1):=0;
\end{rlispverb}
and put the two parts together:
\begin{rlispverb}
ham2:={ham2_l,d,z};
\end{rlispverb}

All that remains to do is to configure the nonlocal variables. Here we have
two distinct operators so we need three nonlocal variables and their names
will be different this time:
\begin{rlispverb}
nloc_var:={{tpsi,w,1},{tpsi,w,2},{tchi,z,1}};
nloc_arg:={{tpsi,w},{tchi,z}};
\end{rlispverb}
Finally we prepare the jet space 
\begin{rlispverb}
dep_var_tot:=cde_weaklynl(
	indep_var,dep_var_equ,loc_arg,nloc_var,total_order);
\end{rlispverb}
and run the computation:
\begin{rlispverb}
sb_res:=schouten_bracket_wnl(
	ham1,ham2,dep_var_equ,loc_arg,nloc_arg);
\end{rlispverb}

We have to plug in all the possible combinations of values $\mu, \lambda = \pm
1$ to obtain the desired result:
\begin{rlispverb}
A:=sub({mu=1,rho=1},sb_res);
B:=sub({mu=-1,rho=1},sb_res);
C:=sub({mu=1,rho=-1},sb_res);
D:=sub({mu=-1,rho=-1},sb_res);
\end{rlispverb}
the result is always zero.

Thus we have proved by a direct computation that previously found operators
$A_1$ and $A_2$ are compatible Hamiltonian operators.

\section{A computation in Maple using \texttt{jacobi}}

In this section we describe another tool which might be used to check if a
weakly nonlocal operator is Hamiltonian and possibly compatible with another
one: the package \texttt{jacobi} \cite{m.20:_weakl_poiss}. Since that package
is written in Maple some readers might find it more convenient, and we thought
that it is worth to explain how to achieve the same result with a different
computer algebra system.

The main structural difference with the computation in the previous section is
that \texttt{jacobi} uses the language of distributions, which is more popular
among Theoretical Physicists.

The plan is the same as before, we need the metrics of both operators to
generate their local parts and the three constants for the nonlocal
part. However the notation is slightly different and that will be the main
point of this section.

For the following computation we also need the external Maple library
\texttt{jacobi.mpl} which can be found on \cite{Gdeq}. The computation itself
is contained in the file\linebreak
\texttt{WDVV-3c-Compat\_Example\_Eta4.mw}. For simplicity, we consider
$\lambda=\mu=1$ for this case.

We start by constructing $A_1$. The metric is declared as follows:
\begin{rlispverb}
g1 := Matrix(N, N);
g1[1, 1] := -u[1, x, 0]^2-u[2, x, 0]^2-4;
\end{rlispverb}
where by \texttt{u[1, x, 0]} we denote in our case $u^1_{0x}$ (with the
convention $u_{0x} = u)$. Other coefficients are defined in a similar
way. After the matrix, we need the Christoffel symbols which can be computed in
Maple or, if we have them available from previous computations, input them as a
three-dimensional field:
\begin{rlispverb}
Gamma1 := Array(1 .. N, 1 .. N, 1 .. N);
Gamma1[1, 1, 1] := -u[1, x, 0];
Gamma1[1, 1, 2] := -u[2, x, 0];
\end{rlispverb}
and so on. Next, we input the two tail vectors \texttt{W1X} and \texttt{W1Y} which in this notation differ only by using a different symbol for the independent variable. The \texttt{W1X} is introduced as follows:
\begin{rlispverb}
W1X := Matrix(N, N)
W1X[1] := u[2, x, 1];
W1X[2] := u[3, x, 1];
W1X[3] := (-u[1, x, 0]*u[2, x, 0]*u[3, x, 1]+
  +u[1, x, 0]*u[2, x, 1]*u[3, x, 0]-
  -u[1, x, 1]*u[2, x, 0]*u[3, x, 0]+
  +u[2, x, 0]^2*u[2, x, 1]+2*u[2, x, 0]*u[3, x, 0]*u[3, x, 1]-
  -u[2, x, 1]*u[3, x, 0]^2+u[2, x, 1])/u[2, x, 0]^2;
\end{rlispverb}
and the same goes for \texttt{W1Y} with $y$ instead of $x$.
The first-order nonlocal operator $A_1$ is then assembled by:
\begin{rlispverb}
A1 := Array(1 .. N, 1 .. N, 0 .. M);
for i to N do 
  for j to N do 
  A1[i, j, 0] := g1[i, j]*delta[x-y, 1]+
    add(Gamma1[i, j, k]*u[k, x, 1], k = 1 .. N)*
    delta[x-y, 0]+W1X[i]*delta[x-y, -1]*W1Y[j]+
    +u[i, x, 1]*delta[x-y, -1]*u[j, y, 1]
  end do 
end do;
\end{rlispverb}

Next for the third-order operator we need its metric $h_{ij}$ as well as
constants $c^{ij}_k$. We input the metric, denoted by \texttt{g3}, as before
and similarly to Christoffel symbols for the previous metric the constants $c^{ij}_k$ are handled as a three-dimensional field:
\begin{rlispverb}
c_hi := Array(1 .. N, 1 .. N, 1 .. N);
c_hi[2, 3, 1] := -1/u[2, x, 0];
c_hi[2, 3, 2] := (u[1, x, 0]-u[3, x, 0])/u[2, x, 0]^2;
c_hi[3, 3, 1] := 1/u[2, x, 0];
\end{rlispverb}
and so on. For the local operator the tail matrix is just a zero matrix:
\begin{rlispverb}
W3 := Matrix(N, N)
\end{rlispverb}
And the operator $A_3$ is assembled by:
\begin{rlispverb}
A3 := Array(1 .. N, 1 .. N, 0 .. M)
for i to N do 
  for j to N do 
    A3[i, j, 0] := g3[i, j]*delta[x-y, 3]
    +add(
      (diff(g3[i, j], u[k, x, 0])+c_hi[i, j, k])*u[k, x, 1], 
      k = 1 .. N)*delta[x-y, 2]
    +add(
      add(
        (diff(c_hi[i, j, h], u[k, x, 0]))*u[k, x, 1]
        *u[h, x, 1], 
        k = 1 .. N)+c_hi[i, j, h]*u[h, x, 2], 
     h = 1 .. N)*delta[x-y, 1] 
  end do 
end do;
\end{rlispverb}
We can then proceed to computing the Schouten bracket itself. We load the
package \texttt{jacobi} by:
\begin{rlispverb}
read `jacobi.mpl`
\end{rlispverb}
The bracket is calculated by the following command and saved into an array
\texttt{T3}:
\begin{rlispverb}
T3 := Array(1 .. N, 1 .. N, 1 .. N);
Schouten_bracket(A3, A3, T3, N, M);
\end{rlispverb}
We can view the result just by printing the contents of \texttt{T3} (further
simplification might be needed):
\begin{rlispverb}
for i to N do 
  for j to N do 
    for k to N do 
      print(T3[i, j, k]) 
    end do 
  end do 
end do;
\end{rlispverb}
The same command is used for computing the Schouten bracket of $[A_1, A_3]$,
proving the compatibility of the two operators:
\begin{rlispverb}
T13 := Array(1 .. N, 1 .. N, 1 .. N);
Schouten_bracket(P1, P3, T13, N, M);
\end{rlispverb}

\section{Large scale computations}

While doing the above computations for $N=3$ is not a time demanding task,
the situation quickly changes for $N=4$, where just finding the third-order
operator usually takes around 5 hours on an average PC. To obtain the results
for $N=5$ in a similar way one would have to have an access to a huge amount of
computational resources.

In this section, we describe the way in which we found the third-order
operators in the case $N=5$ (results in \cite{VV21}).  Since
we did not even try to find a first-order operator when $N=5$, we will not deal
with related problems. Only partial results are known in the case $N=4$
\cite{PV15}.

There are two ways how to approach this problem. First, we could find a
powerful enough machine to handle this computation. However, we are talking
about hundreds of GB of RAM memory and possibly a week of computing time.

The other option is to optimize the computation itself, so that it not so
resource-demanding. It can be done in a few ways, but the most effective at the
beginning is to reduce the number of equations we try to solve at once.

As both approaches were used to obtain the results for $N=5$ let us explore
those below in more detail. All files used below are in a separate folder \linebreak \texttt{\textbackslash WDVV-5c-Large\_scale}.

\subsection{Using a supercomputer}

If you have an access to a supercomputer it could be beneficial to try to use a
rather ``brute force'' approach. Given the fact we cannot say for sure how many
equations need to be solved in order to obtain the solution, it is by far the
best first try.

To obtain the result in case of $N=5$ for a canonical choice of $\eta$ we have
used the whole super-computing grid provided by a virtual organization
\emph{CESNET -- MetaCentrum}. The command to start the computation on the grid
can be found in the file \texttt{starter.sh}.

It is also important to mention that we need to provide the file with the
generated system in the conservative form as the algorithm has not been
transferred into Maple yet. The algorithm itself is contained inside the file
\texttt{w10\_hydro\_system\_gen.red} whose output file
(\texttt{w10\_eta2\_eq.red}) is then loaded into Maple. Finding the
quasi-linear system in the conservative form using an algorithm was described
in detail in \cite{VV21} and is an interesting computational problem. However,
since the program literally follows the steps described there, we will omit
repeating the details. Note that some syntax adjustments are required before
importing.

After we obtain the result from a remote computation it is important to have an
independent check. We will use Reduce, file
\texttt{w10\_ham1\_eta2\_check.red}. We initialize as usual:
\begin{rlispverb}
load_package cde;

nc:=10;

indep_var:={x};
dep_var:=for i:=1:10 collect mkid(a,i);
total_order:=6;
resname:="w10_ham1_eta_check_res.red";

cde({indep_var,dep_var,{},total_order},{});

in "w10_eta1_eq_transformed.red";

cap_v:=for each el in cons_laws_system collect second el;

in "w10_eta_gl3.red";
\end{rlispverb}
where in the files \texttt{w10\textunderscore eta2\textunderscore eq.red} and
\texttt{w10\textunderscore eta\textunderscore gl3.red} we saved the general
WDVV system and the result of the computation -- the metric $h_{ij}$
respectively which again needs to be transformed into the proper syntax.

Then we need to confirm that the equations \eqref{eq:3rdOrdCon} holds for the
metric $h_{ij}$ already in place (in form of the constants $c_{ijk}$) to
confirm the compatibility with the first-order system. Lastly we need to check
the Hamiltonian condition which we assemble by:
\begin{rlispverb}
gu3:=gl3**(-1);
for k:=1:nc do
 for l:=1:nc do
  for m:=1:nc do
   for n:=1:nc do
  << if
      df(c_lo(m,n,k),part(dep_var,l)) + (
	for i:=1:nc sum (
	  for j:=1:nc sum ( 
	    gu3(i,j)*c_lo(i,m,l)*c_lo(j,n,k)
	    ))
	) neq 0 then rederr "Error - not Hamiltonian!"
  >>
\end{rlispverb}
which also passes in our case and confirms the result obtained by a
super-computation.

\subsection{Simplifying the computation}

If we look into the algorithm we can quickly realize that it is massively
inefficient in the sense of computational resources. The unknowns are exactly
the coefficients of the quadratic polynomials in the metric of the third-order
operator. In the case of $N=5$ we need to find $1540$ unknown constants.

On the other hand, the conditions \eqref{eq:3rdOrdCon} generate 2100 equations
but those equations are polynomial with respect to the field variables. We need
to collect the coefficients of each monomial out. There will be an immense
amount of those collected equations. Just for comparison, in the case $N=4$ we
have 231 unknown coefficients, slightly more than 450 equations polynomial in
the field variables and we generate around 350,000 of such linear
equations.

It is now obvious that solving all of the collected equations is completely
unnecessary and in this case a waste of resources. We could try to make an
algorithm that will solve only a few of the equations which are polynomial in
the field variables at once, plug the result into the rest and repeat the
process batch by batch until all the equations are solved. This process could
work, however, it is very hard to predict how much resources it will consume
and what is the optimal size of the mentioned batch. The two extreme cases
being the batch is too small and the number of iterations will project into the
time of the computation or the batch being too big literally clogging the
memory and maybe never finish.

There is certainly enough space to optimize the whole process of solving those
massively over-determined system of linear equations. However, by experiment we
have found out that in our case there are around 60 equations out of the
original 2100 polynomial ones needed to completely solve the posed
problem. They originate from the system \eqref{eq:15} below.

Also, saving partial results in the whole process is an absolute must. While
the setup of the problem does not consume much time, relative to solving it, it
is much easier to plug in a partial result from earlier and continue the
computation.

More specifically, instead of solving the whole set of equations
\eqref{eq:3rdOrdCon} we can simply restrict the system to
\begin{equation}\label{eq:15}
h_{ik}V^{k}_{j}=h_{jk}V^k_{i}, \qquad i,j,k=1,\dots,4. 
\end{equation}
This computation takes around 8 hours on an average PC and the result is
checked in the same way as in the previous case.

\section{Conclusions}
\label{sec:conclusions}

In this paper we described the computations that we performed in \cite{VV21} in
order to support the conjecture that all WDVV equations are endowed with
bi-Hamiltonian formalism. These are symbolic software calculations that are
nontrivial under both the viewpoint of the mathematical algorithm and the
viewpoint of the symbolic programming that is needed to achieve the goal.

However, we think that further computational goals might be achieved to support
and broaden the above conjecture.

\begin{itemize}
\item First of all, in the $N=4$ case the bi-Hamiltonian property has been
  proved only for one of the two canonical forms of $\eta$, $\eta^{(1)}$. This
  was a result found in \cite{PV15}. For the system arising from the second
  canonical form $\eta^{(2)}$ we found a third-order operator but we haven't
  even tried to find a first-order operator. It should be said that the method
  that we used in the case $N=3$, coming from a characterization in
  \cite{bogoyavlenskij96:_neces_hamil}, is not applicable in higher dimensions,
  hence the calculation might be a real challenge. Of course, the case $N=5$ is
  even harder.

  A way to find a first-order operator might be to restrict the search to
  matrices $g^{ij}$ of rational functions of a certain degree. This common
  pattern can be observed in our examples (see also \cite{VV21}) and might
  greatly simplify the search.

  More generally, it should be possible to program a \texttt{cde} or
  \texttt{jacobi} computation of all first-order nonlocal operators that are
  compatible with a given third-order operator. This is a task for a future
  research project.
\item Then, there is a well-known generalization of the WDVV equations, the
  oriented associativity equation\cite{LosevManin04}. These equations and the
  associated geometric structures are called F-manifolds with compatible flat
  structures \cite{Manin05}, or simply flat F-manifolds. They share many
  properties with WDVV equations and Dubrovin-Frobenius manifolds including the
  existence of an associated integrable dispersive hierarchy (see
  \cite{ABLR20:_semis_f} for details).

  We observe, in particular, that the oriented associativity equation has an
  infinite hierarchy of nonlocal symmetries \cite{Sergyeyev_09}, a first-order
  local Hamiltonian operator of the same type as $A_1$ \cite{pavlov14:_orien}
  and a third-order \emph{nonlocal} Hamiltonian operator which is the
  straightforward generalization of $A_2$
  \cite{casati19:_hamil,m.v.19:_bi_hamil_orien_assoc}.

  Until now, the results on Hamiltonian operators are known only for one of the
  simplest cases of oriented associativity equation; more calculations are
  needed in order to support a conjecture on the bi-Hamiltonianity of the
  F-manifold equation. Indeed, the compatibility of $A_1$ and $A_2$ is still an
  open question even in the simplest case, see
  \cite{m.v.19:_bi_hamil_orien_assoc} for a discussion.
\end{itemize}
The above problems are on our schedule.

\begin{acknowledgements}
  R.V. would like to thank P. Lorenzoni and A. Sergyeyev for useful comments
  and suggestions.

  Computational resources were supplied by the project ``e-Infrastruktura CZ''
  (e-INFRA CZ LM2018140) supported by the Ministry of Education, Youth and
  Sports of the Czech Republic.

  The research of J.V. was supported in part by the Specific Research grant
  SGS/13/2020 of the Silesian University in Opava, Czechia.

  The research of R.V. has been funded by the Dept. of Mathematics and Physics
  ``E. De Giorgi'' of the Universit\`a del Salento, Istituto Naz. di Fisica
  Nucleare IS-CSN4 \emph{Mathematical Methods of Nonlinear Physics}, GNFM of
  Istituto Nazionale di Alta Matematica.
\end{acknowledgements}

The authors declare that they have no conflict of interest.

\providecommand{\cprime}{\/{\mathsurround=0pt$'$}}

\end{document}